\documentclass[a4paper]{article}

\usepackage{INTERSPEECH2022,url}
\usepackage{subcaption}
\usepackage{xcolor}
 \usepackage{array,multirow,graphicx}

\definecolor{cornflowerblue}{rgb}{0.39, 0.58, 0.93}
\definecolor{palegoldenrod}{rgb}{0.93, 0.91, 0.67}
\title{Study on the Fairness of Speaker Verification Systems on \\ Underrepresented Accents in English \vspace{-0.14cm}}
\name{Mariel Estevez, Luciana Ferrer}
\address{\vspace{-0.2cm}
  Instituto de Investigación en Ciencias de la Computación,
  UBA-CONICET, Argentina}
\email{maestevez@icc.fcen.uba.ar, lferrer@dc.uba.ar}

\begin{document}

\maketitle
\begin{abstract}
\vspace{-0.1cm}

Speaker verification (SV) systems are currently being used to make sensitive decisions like giving access to bank accounts or deciding whether the voice of a suspect coincides with that of the perpetrator of a crime. Ensuring that these systems are fair and do not disfavor any particular group is crucial. 
In this work, we analyze the performance of several state-of-the-art SV systems across groups defined by the accent of the speakers when speaking English. To this end, we curated a new dataset based on the VoxCeleb corpus where we carefully selected samples from speakers with accents from different countries. We use this dataset to evaluate system performance  for several SV systems trained with VoxCeleb data. We show that, while discrimination performance is reasonably robust across accent groups, calibration performance degrades dramatically on some accents that are not well represented in the training data. Finally, we show that a simple data balancing approach mitigates this undesirable bias, being particularly effective when applied to our recently-proposed discriminative condition-aware backend.
  
\end{abstract}

\section{Introduction}
\vspace{-0.1cm}

In recent years, a large number of works have been published on the issue of fairness of AI systems. The studies seek to understand whether sensitive attributes like gender or age affect the system's performance, potentially resulting in unfair behavior toward a group of individuals \cite{barocas-hardt-narayanan}. Most publications on fairness focus on tasks like face recognition \cite{face_recognition,face_recognition2}, natural language processing \cite{NLP,NLP2}, and automatic speech recognition \cite{ASR,ASR2}. In this work, we focus on the speaker verification (SV) task, where the goal is to decide whether two audio samples (or two sets of samples) belong to the same or to different speakers. SV systems are used for a large variety of applications, such as accessing bank accounts, searching for individuals of interest within large audio datasets, or verifying whether a certain suspect is the same person that was recorded while committing a crime -- an application called forensic voice comparison \cite{Morrison_Thompson_2017,danieltesis}. In some of these applications, mistakes can have severe consequences, like letting an impostor withdraw money from somebody else's account, or declaring an innocent person guilty of a crime. 

State-of-the-art (SOTA) SV systems have impressive performance when acoustic conditions are relatively clean and the subjects' gender, age, language, accent, as well as other characteristics like the speaking style, are well represented in the data used for training the system. On the other hand, when the attributes of the test subjects differ from those seen in the majority of the training speakers, even in relatively clean conditions the performance --most notably, calibration-- can degrade significantly \cite{ferrer2021}. Hence, assessment of the fairness of SV systems across minority groups is an important issue. 
Notably, to date very little work has been done on the evaluation and mitigation of bias in SV systems. Among the few works on this topic, a recent publication \cite{fenu2020} showed a disparity between the system performance for different age and gender groups. This work uses the FairVoice dataset \cite{fenu2020}, curated by the authors, which is composed of short clean read sentences from the Mozilla Common Voice collection \cite{common}. Another work in this area \cite{sveva} studies the disparities in an SV system on VoxCeleb data \cite{Nagrani19}, a corpus widely used for SV system training and evaluation. In this work, authors group the data by nationality and gender of the speaker and show that models perform worse on females than males for most nationalities. 
Finally, in \cite{stolcke} authors study the effect in SV discrimination performance of gender imbalance in the training data and propose an approach to mitigate the disparity that results from this imbalance.

These works indicate that modern SV systems suffer from bias on certain groups. We believe this problem deserves further attention. In particular, no prior work has looked into the effect that a speaker's accent has on the system. While nationality of a speaker can be considered a proxy for the accent they have when speaking English, many speakers from non English-speaking countries have near-native accents in English. Further, some percent of the samples in VoxCeleb are not in English. These issues, as well as the small number of speakers for some nationalities, might explain why the conclusions in \cite{sveva} are inconsistent when focusing on nationality. Hence, in this work, we aim to directly study the effect of accent on SV system performance.
The contributions of this paper are: 1) the creation of a new dataset derived from VoxCeleb data with curated accent information; (2) the detailed analysis of several SOTA SV systems' performance on this dataset, focusing on calibration-sensitive metrics; and (3) the exploration of a balancing approach to mitigate the bias found to occur across accent groups.  Our results indicate that, as expected, calibration performance on most non-native English accent groups degrades severely with respect to native English accent groups for all tested SV systems. The proposed balancing approach greatly reduces the miss-calibration for all accent groups and systems, reaching almost perfect calibration for the system that uses a recently proposed discriminative backend with condition-aware calibration \cite{ferrer2021}.

\vspace{-0.1cm}
\section{The VoxAccent dataset}
\vspace{-0.1cm}

For this work, we curated a new dataset using, as in \cite{sveva}, the VoxCeleb corpus  as basis. VoxCeleb \cite{Nagrani19} contains speech collected from YouTube videos, including speech from over 7,000 public figures from many countries speaking mostly English, though other languages are also present. We obtained the nationality of the speakers through an automated search in Wikipedia. Then, we randomly chose 35 speakers from each of the following nationalities: USA, UK, Australia, Canada, Germany, France, Italy, Spain, and India.
For the non-English speaking countries, we listened to samples from each of these speakers to verify that they had a noticeable non-native accent when speaking English, discarding those that sounded very close to native English speakers. We also discarded any speakers that were mentioned in Wikipedia to have grown up in more than one country.

After this selection, some groups had much fewer females than males. Hence, for this initial version of the dataset, we decided to select only male speakers.  For future versions of the dataset, we will go back to the first step and attempt to select a similar number of female speakers for each group. For each of the selected speakers, we chose up to four recordings among those spoken in English. Further, we discarded very degraded or noisy audios, in an effort to achieve somewhat homogeneous audio quality across different accent groups to avoid confusing a bias due to audio quality with bias due to accent. A few speakers were left with a single audio file after this selection, in which case we discarded the speaker. After the full selection process, the resulting dataset has between 17 and 19 speakers for each accent group, and a total of 159 speakers. Finally, to ensure homogeneity in terms of speech duration, we cut the original waveforms into 4 (potentially overlapping) chunks containing approximately 16 seconds of speech, spreading the starting times of the chunks uniformly across the file. We created trials pairing all chunks from all speakers against all other chunks, except for those chunks that come from the same original audio file. The number of same-speaker trials for each country varies between 1400 and 1800, while the number of different-speaker trials varies between 30000 and 43000. 
The full specification of the dataset, as well as the code needed to run the experiments in this paper, can be requested to the authors.

\section{Speaker verification systems}
\vspace{-0.1cm}

Most current speaker verification systems are composed of a cascade of multiple stages. First, frame-level acoustic features that represent the short-time content of the signal are extracted. These variable-length sequences of features are input to a deep neural network (DNN) which is trained to optimize speaker classification performance on the training dataset. The DNN uses a temporal pooling layer so that it can represent the speaker information in the variable-length input as new features of fixed dimension that are termed the \emph{speaker embeddings} (or embeddings, for short), which are extracted from a hidden layer in the DNN after the temporal pooling. The embeddings are then typically transformed using linear discriminant analysis (LDA), and further normalized. Next, probabilistic  linear discriminant analysis (PLDA) \cite{Ioffe:2006,prince:plda} is used to obtain scores for each speaker verification trial composed of an enrollment and a test sample to be compared. Finally, if required, a calibration stage can be included to convert the scores produced by PLDA into likelihood ratios. More details and references on the standard SV pipeline can be found, for example, in \cite{ferrer2021}.

In this work, we analyze the performance of 12 different speaker verification systems obtained through the combination of two different speaker embedding extraction models (XVECT and ECAPA), three different backend approaches (PLDA, DPLDA and DCAPLDA) and two approaches for balancing the training data (no balance, balancing by nationality). The following sections explain these different components.

\vspace{-0.2cm}
\subsection{Speaker embedding extractors}
\vspace{-0.1cm}
We use two different pre-trained embedding extractors from the SpeechBrain toolkit \cite{speechbrain}. The models were trained by the authors of the toolkit using VoxCeleb data. While the training data includes our test data, we believe this should not affect our general conclusions since the model is trained with over 250,000 audio samples from more than 7000 speakers, making it unlikely that results would change significantly after discarding our test data composed of 159 speakers. Further, while results may be somewhat optimistic, the observed trends of biases should still hold.  Nevertheless, in the near future we plan to retrain the embedding extractors after discarding all data from our test speakers. Importantly, this issue only applies to the embedding extractors, the backends are trained without including any of the test speakers' data.

The two models we use correspond to the now traditional X-vector architecture \cite{xvector,xvectorlink} (XVECT for short), and to the more recent ECAPA-TDNN architecture \cite{ecapa,ecapalink} (ECAPA for short). We do not include details on the architectures due to lack of space, but rather highlight the differences between the two models, which are the input features (24 mel filterbanks for XVECT, 80 for ECAPA), the number of parameters (ECAPA has over 4 times more parameters than XVECT), the training loss (cross-entropy for XVECT, additive angular loss for ECAPA), and the size of the embeddings (512 for XVECT, 192 for ECAPA). The authors report 3.2\% EER on a cleaned Voxceleb1 test set for XVECT and 0.69\% for ECAPA.

While the SpeechBrain toolkit does not filter out non-speech frames before feeding them to the embedding extractor, in a separate yet-unpublished study we found that applying a state-of-the-art speech activity detector (SAD) gave better results on a wide range of conditions. Hence, for this work we filter out non-speech frames before embedding extraction.
The speech regions were computed with SRI International’s Speech Detection model described in \cite{ferrer2021} and used for this work with their permission.

\vspace{-0.2cm}
\subsection{Backends}
\vspace{-0.1cm}
We use three different backends: the standard PLDA backend, and two discriminative backends recently proposed in \cite{ferrer2021}, DPLDA and DCAPLDA. DPLDA takes the same functional form as the standard PLDA backend, including the LDA, length normalization, PLDA and calibration stages, with the difference that all the parameters are trained jointly and discriminatively after being initialized with the values obtained with the standard approach. The discriminative fine-tuning of the parameters is made using stochastic gradient descent of the binary cross-entropy. A more complex version of this backend modifies the calibration stage making its parameters depend on an estimate of the condition and speech duration of the two signals involved in a trial. The estimate of the condition is given by a low-dimensional embedding that is obtained from the original speaker embedding with a small DNN, which is learned jointly with the rest of the model's parameters. The speech duration is obtained with the SAD system mentioned above. In \cite{ferrer2021} we show that this method, which we call DCAPLDA (discriminative condition-aware PLDA), leads to robust performance across a wide variety of conditions, when the model is trained with many different datasets balanced to give each dataset a similar number of terms in the training loss. We found that this approach allowed us to leverage the large amount of training data available much more effectively than simple PLDA or DPLDA since it is able to dynamically and automatically adapt the output scores to the condition of the input samples.

The backends are trained with the same process described in \cite{ferrer2021}, using the subset of the Voxceleb 1 and 2 datasets obtained by discarding the speakers included in VoxAccent.  Further, we randomly chunk the training data to achieve an approximately uniform distribution between 4 and 240 seconds in the log domain. The standard test set from Voxceleb2 (VOX2) chunked to durations between 4 and 16 seconds, is used for selection of the best epoch and seed for the DPLDA and DCAPLDA approaches. For backend training and evaluation we use the DCAPLDA toolkit available at \url{https://github.com/luferrer/DCAPLDA}. 

\vspace{-0.2cm}
\subsection{Training data balancing}
\vspace{-0.1cm}
When the training data is imbalanced in terms of one or more characteristics (gender, acoustic conditions, or, as in our case of interest, accent), training the models naively without any effort to balance out the influence of the different groups is, in many cases, similar to simply training the model with only the majority groups. Minority groups are disregarded during training since they have little influence on the training loss, resulting on suboptimal performance on those groups \cite{ferrer2021}.

The simplest approach for balancing the training data is to downsample the majority groups to the size of the minority groups. This approach, though, is suboptimal since it discard potentially valuable data. In our prior work \cite{ferrer2021} we proposed to use simple balancing approaches during backend training that do not require discarding any data. In the case of PLDA, each sample is assigned a weight proportional to the number of samples in its group. These weights are then used during the estimation of the LDA and PLDA parameters, balancing out the influence of each group. In the case of DPLDA and DCAPLDA, the balance is achieved by creating batches with the same number of samples for every group. For details on these balancing approaches, see \cite{ferrer2021}.

In this work, we use these approaches to balance out the different accents during training. Since accent information is not available for the training data, we use the automatically obtained nationality information as a proxy. Further, since the training data includes a long tail of many nationalities with very few speakers, we create individual groups only for the nationalities with at least 100 speakers in training (which include all nine of the nationalities in VoxAccent) and group all others into a single ``other nationality'' category.

\vspace{-0.1cm}
\section{Calibration and fairness metrics}
\vspace{-0.1cm}

\label{sec:metrics}
In our prior work we have observed that calibration performance of SV systems degrades much more rapidly and severely than their discrimination performance when testing on conditions that are underrepresented in the training data \cite{ferrer2021,Ferrer:aslp18}. That is, while the separation between classes could still be reasonable on a certain minority group, the scores may be shifted or warped resulting in poor calibration making them uninterpretable and suboptimal for decision making based on thresholds given by Bayes decision theory or optimized globally on some equally-biased dataset.  For this reason, we propose to use a calibration-sensitive metric to measure performance. A widely used calibration-sensitive metric for the SV task is the Cllr \cite{van2007introduction}, which is given by the average of the cross-entropy (CE) loss for the two classes (same-speaker and different-speaker).
The Cllr measures the overall quality of the scores, including discrimination and calibration aspects, across all possible operating points. Further, the Cllr value on a certain test set obtained after transforming the scores with the best monotonic transformation can be used to obtain a measure of the discrimination performance of the system. In this paper, we restrict the transformation to be affine, given the small number of samples in each accent group. The difference between the actual and the minimum Cllr, usually called \emph{calibration loss}, indicates how much of the Cllr is due to poor calibration as opposed to poor discrimination. For an introductory explanation on optimal Bayes decisions, calibration, and Cllr, please see \cite{van2007introduction}. 

In this work, we use a slightly modified version of the Cllr, defined in  Equation (14) in \cite{ferrer2021}, where rather than taking an average of the CE for the two classes, we weight the term for the same-speaker class by a parameter $\pi$ and the term for the different-speaker class by $1-\pi$. In this work $\pi$ is set to 0.05, which is approximately the prior for the same-speaker class in each of our test groups. Using this value for $\pi$ gives a more robust metric than the standard Cllr where the average CE on the few same-speaker samples is weighted equally to the average CE on the many more different-speaker samples.
Further, this value of $\pi$ effectively penalizes errors on the different-speaker samples more than errors on same-speaker samples, which is the right approach when false alarms (different-speaker trials labelled as same-speaker trials) are more costly than misses (same-speaker trials labelled as different-speaker trials), as is the case of forensic voice comparison. In addition, since, as we will show in the next section, the bias on the minority groups is such that the false alarm rate dramatically increases with respect to majority groups, we believe it is important to choose a metric that highlights this issue by penalizing false alarms more than misses.
Finally, given that each group in the curated dataset is relatively small, we use the boostrap method \cite{poh:confidence:07} to obtain confidence intervals on the Cllr.

We also compute the Fairness Discrepancy Rate (FDR) as defined in Equation (6) in \cite{FDR}. This metric is given by a weighted average of two terms: the largest distance between false alarm rates for any two groups, and the largest distance between miss rates. The weight, called $\alpha$ is equivalent to $1-\pi$ in the Cllr definition, so we set it to 0.95. Accordingly, we choose the decision threshold needed to compute the FDR as the Bayes decision threshold for a cost function where the miss rate is weighted by $\pi$ and the false alarm rate is weighted by $1-\pi$. The Bayes threshold is the one that optimizes the cost function when scores are well calibrated (see, e.g., \cite{van2007introduction}). For these weights, its value is 2.94. Note that FDR, as the Cllr, is also calibration-sensitive, though, unlike Cllr, it focuses on a single operating point. A value of FDR of 1.0 corresponds to a perfectly fair system for the selected threshold.

\section{Results and discussion}
\vspace{-0.1cm}

\begin{figure*}[t]
\begin{subfigure}{.75\textwidth}
    \centering
    \includegraphics[width=0.99\textwidth]{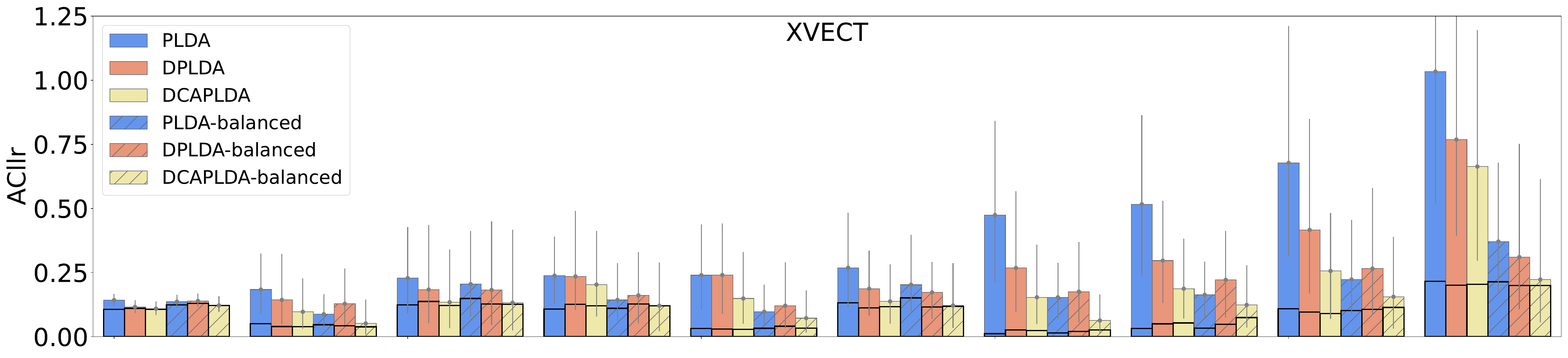}
\end{subfigure}%
\begin{subfigure}{.25\textwidth}
    \centering
    \vspace{-.2cm}
    \includegraphics[width=0.58\textwidth]{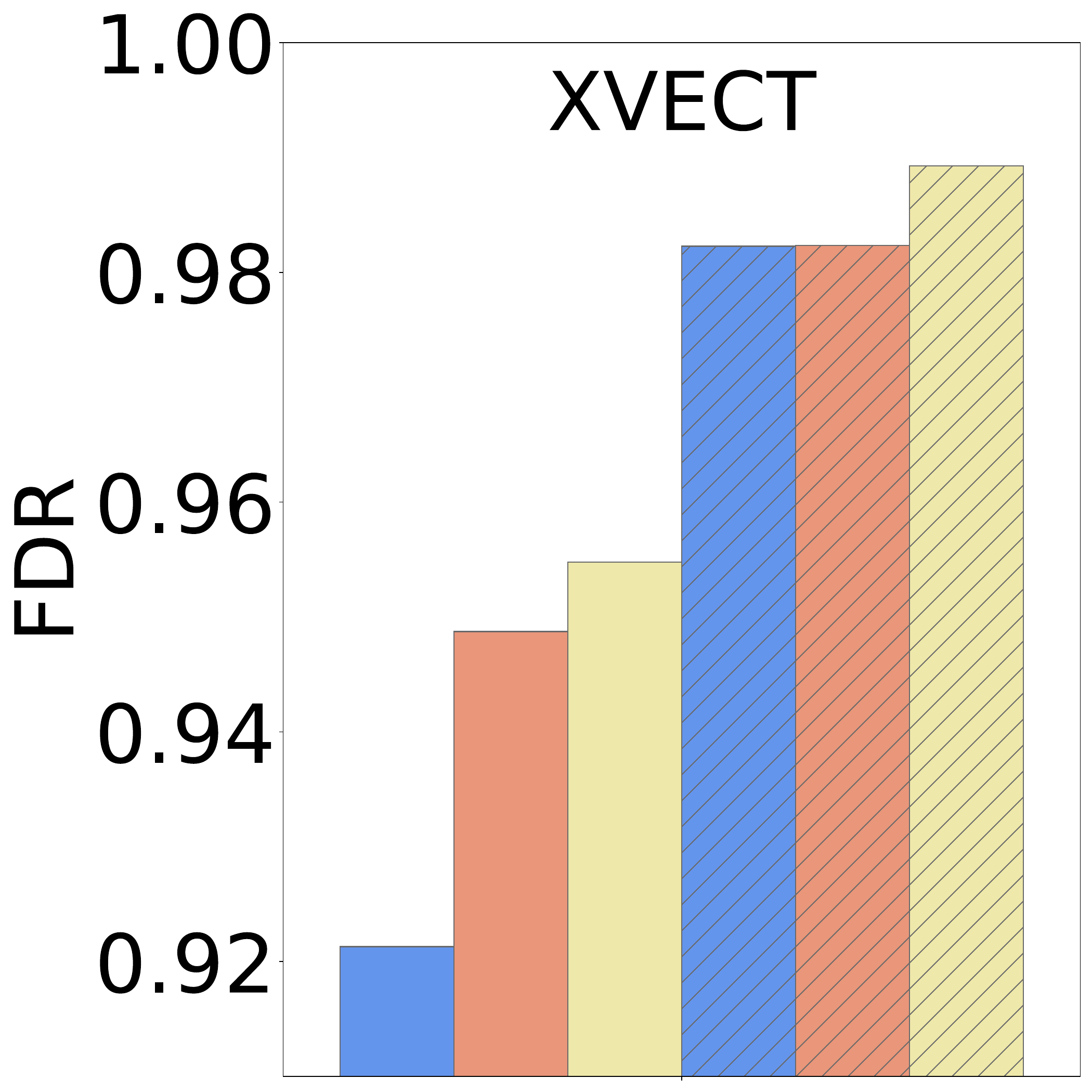}
\end{subfigure}%

\begin{subfigure}{.75\textwidth}
    \centering
    \includegraphics[width=0.99\textwidth]{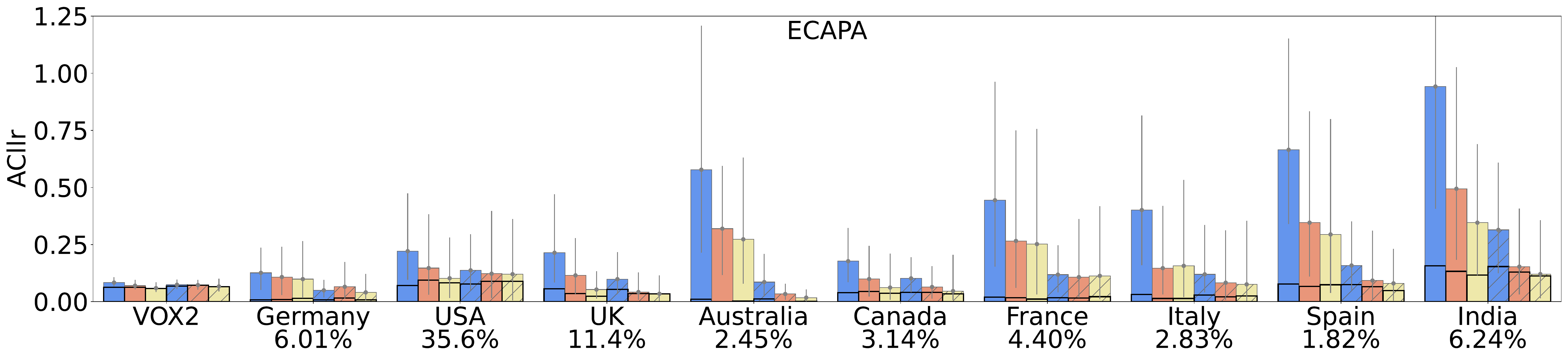}
\end{subfigure}%
\begin{subfigure}{.25\textwidth}
    \centering
    \vspace{-.3cm}
    \includegraphics[width=0.58\textwidth]{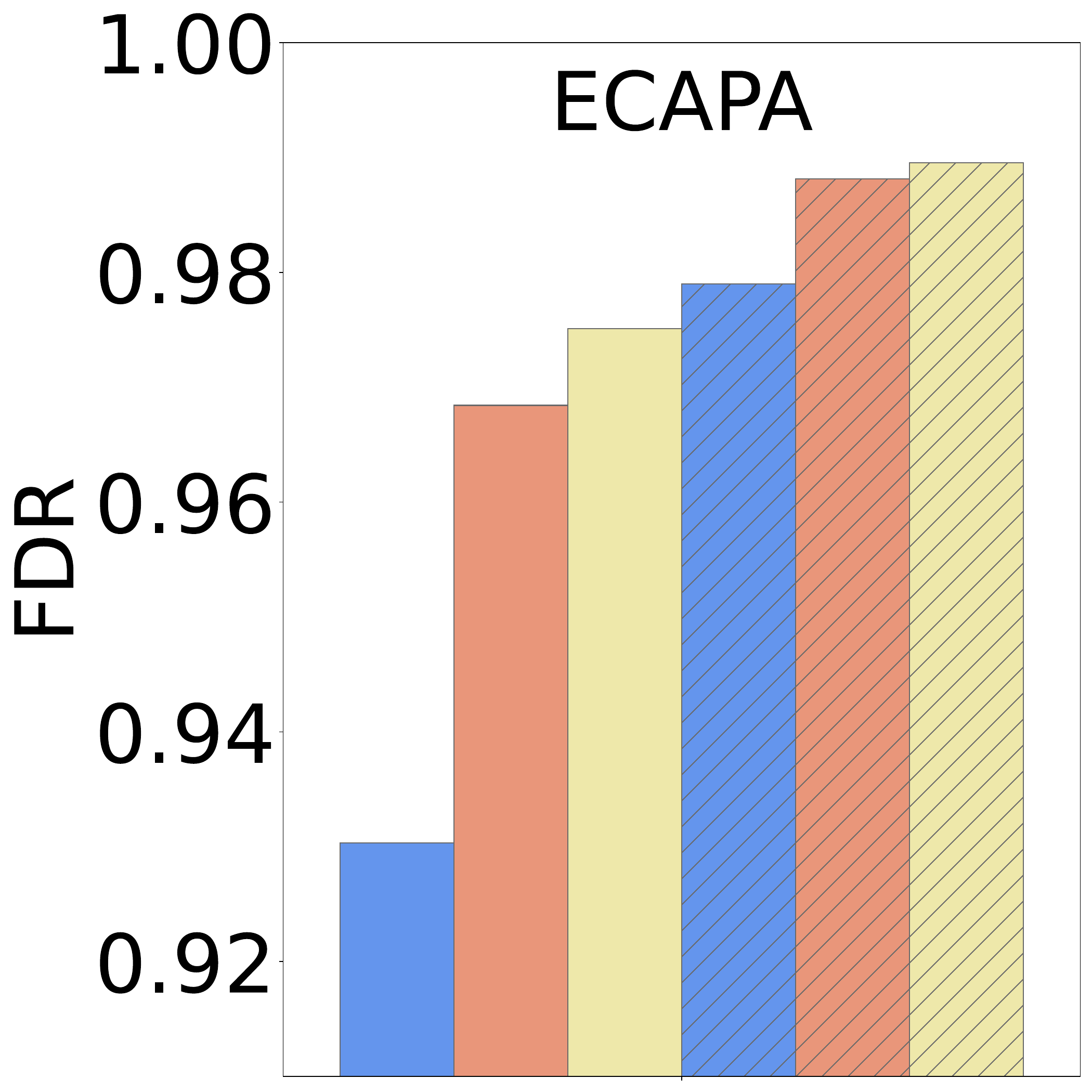}
\end{subfigure}%
\vspace{-0.2cm}
\caption{On the left:
Actual Cllr (bar height) and minimum Cllr (black lines inside bars) for all systems and all accent groups. Countries are sorted by their PLDA performance without balancing for the XVECT embeddings.  The vertical lines correspond to the confidence intervals. The numbers under the country names are the percentage of training data corresponding to that country. On the right: FDR for each of the six systems (higher is better in this case).}
\label{fig:barplots}
\end{figure*}

Figure \ref{fig:barplots} shows the results for all systems tested in this work over all accent groups, as well as on the VOX2 test set (chunked to 16 seconds, as for VoxAccent). We can see that, while there is a correlation between the performance of the baseline system that uses a PLDA backend without balancing and the frequency of the country in the training data (listed under each country's name), the relationship is not direct. Germany and Canada are infrequent but they have a Cllr comparable to the most frequent countries, USA and UK. We believe this could be explained by the fact that the German and Canadian accents share a lot of phonetic aspects with USA and UK accents. On the other hand, for France, Italy, Spain and India, which are both infrequent and have a phonetic inventory that has many differences with that of USA and UK, the performance degrades dramatically, as expected.  Australia is a unusual case, behaving like Canada and Germany for the XVECT embeddings, but like the other minority countries for ECAPA embeddings. This may be due to the ECAPA model being able to overfit more heavily to the frequent data than the XVECT model. 

Comparing performance across systems, we can see that, without balancing, DCAPLDA gives gains over PLDA and DPLDA for most countries and both embeddings. The largest gains, though, are obtained with the balancing approach which reduces Cllr dramatically for most countries and both types of embeddings, reaching close to zero calibration loss in most cases,  particularly so for the DCAPLDA backend. Importantly, performance on the majority groups and on VOX2 is not degraded by the data balancing approaches. Note that, as mentioned above, all countries in the test groups have enough samples in training to get their own group for the balancing approach. We tested the performance of the DCAPLDA-balanced backend after discarding the India accent group from training and the performance on the India test set dropped to the same level as without balancing, indicating that the model is not able to generalize well across accent groups. We hypothesize that if the training data contained a large number of different accents with enough speakers, generalization could be improved. This, unfortunately, cannot be tested with this dataset. 

Interestingly, Figure \ref{fig:barplots} shows that the discrimination performance, measured by the minimum Cllr, is less dependent on the accent than the total performance measured by the actual Cllr. While countries differ in terms of discrimination performance, the differences do not seem to be due to the accent but rather to the particular set of speakers we selected in each case. This hypothesis is supported by the fact that balancing does not have a significant effect on discrimination performance, and that the confidence intervals for the minimum Cllr (not shown) are extremely wide, overlapping completely with each other across countries. Hence, we believe that most of the difference in discrimination performance would disappear given a much larger sample of speaker for each country. We hope to be able to increase the  size of the dataset in the near future.

The right plot in Figure \ref{fig:barplots} shows the FDR values for all systems. The FDR can be seen as a summary of the fairness of a system for the selected threshold. We can see that DCAPLDA gives gains over the other two backend approaches both with and without data balancing. After data balancing, the DCAPLDA backend reaches FDR values of 0.989 and 0.990 for XVECT and ECAPA, respectively.

\begin{figure}[t]
    \vspace{-.2cm}
\begin{subfigure}[b]{0.43\textwidth}
    \centering
    \hspace*{-3.5cm}
    \vspace{-.2cm}
    \includegraphics[width=0.6\columnwidth]{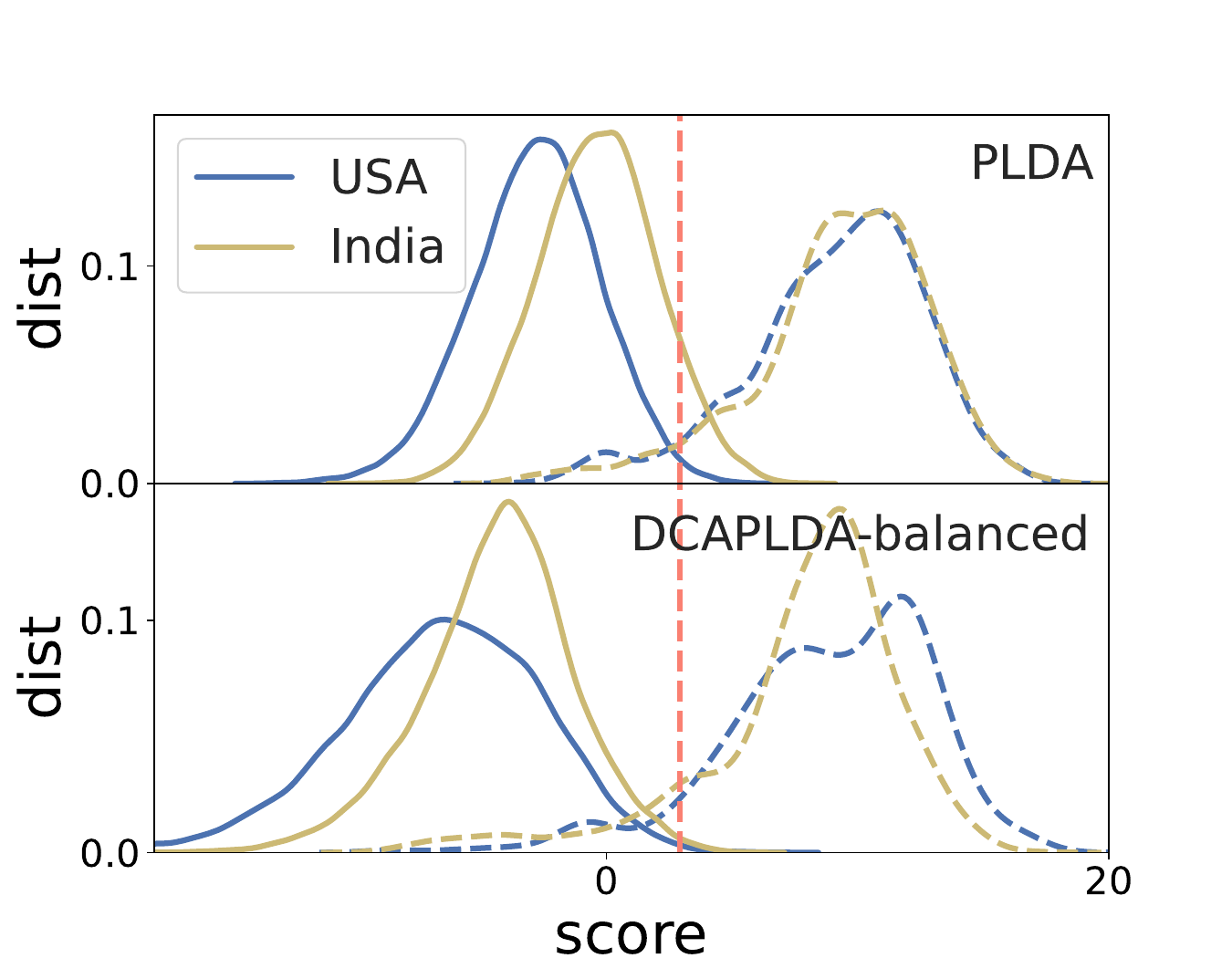}
    
\end{subfigure}
\setlength{\tabcolsep}{2.0pt}
\begin{subtable}[b]{0.03\textwidth}
\footnotesize
  \centering
  \hspace*{-4.0cm}
  \vspace{.6cm}
  \scalebox{0.9}{
  \begin{tabular}{c c c c }
  
   & Group & $P_{\textrm{fa}}$ & $P_{\textrm{miss}}$  \\
    \toprule
  \multirow{2}{*}{PLDA} 
  &  USA & 0.91 & 5.69  \\
    & India & 8.63 & 5.58  \\
    \midrule 
  DCAPLDA  & USA & 0.26 & 7.64 \\
  balanced  & India  & 0.51 & 10.8 \\
  \end{tabular}}
\end{subtable}
\caption{Score distribution for XVECT embeddings, for two backends, for USA and India groups. Solid and dashed lines correspond to different- and same-speaker trials, respectively. The false alarm and miss rates are computed for the threshold used to compute FDR, indicated by the dashed vertical line.}
\label{fig:densities}
\end{figure}

Finally, Figure \ref{fig:densities} shows the score distributions for two accent groups for the baseline system and for the system with the best FDR. We can see that, for the baseline, the distribution of different-speaker scores for India shifts to the right with respect to the distribution for USA, explaining the large calibration loss observed for that system on that accent group. As a consequence of this shift, the false alarm rate for the same threshold is much larger for India than for USA (see table beside the plots). If such system was used for a forensic voice comparison it would have a much higher chance (about 10 times higher) of suggesting that the suspect is guilty of the crime when the suspect has an Indian accent than when they have a US accent. The bottom plot shows the score distributions for the DCAPLDA backend trained with the balancing approach. We can see that the shift in the different-speaker distribution is reduced. The false alarm rate is now similar for both countries, though both error rates are larger for India than for USA, since discrimination performance is worse in the former. As noted above, we believe this difference in discrimination performance may be a random effect due to the specific set of speakers selected for each country.

\vspace{-0.2cm}
\section{Conclusions}
\vspace{-0.1cm}
In this work we studied the performance of various SV systems across different groups defined by the speakers' accent in English. We showed that performance measured with a calibration-sensitive metric is significantly worse on some accents that are not well represented in the training data than on majority accents. This is due to miscalibration of the scores for the different-speaker trials from those accent groups, which are shifted toward the right causing a striking increase in false alarm rate. 
We showed that applying simple data balancing approaches during training greatly reduces miscalibration in the affected groups, without degrading performance on the majority groups. 


\bibliographystyle{IEEEtran}

\bibliography{mybib}

\end{document}